Climatic implications of a rapid wind/solar transition

Peter D. Schwartzman[1*], David W. Schwartzman[2], Xiaochun Zhang[3*]

[1]Department of Environmental Studies, Knox College, Galesburg, Illinois 61401, USA.

[2]Department of Biology (Emeritus), Howard University, Washington, DC 20059, USA.

[3]Department of Global Ecology, Carnegie Institute for Science, Stanford University, Stanford, California 94305, USA.

*e-mail: pschwart@knox.edu; xczhang@carnegiescience.edu

Abstract

A transition to a fully global renewable energy infrastructure is potentially possible in no more than a few decades, even using current wind/solar technologies. We demonstrate that at its completion this transition would terminate anthropogenic carbon emissions to the atmosphere derived from energy consumption in roughly 25 years as well as double current global energy production. This result would provide all human energy needs worldwide and additional energy required for climate adaptation as well as carbon sequestration from the atmosphere to bring down the atmospheric carbon dioxide ($CO_2$) concentration to safer levels. The implementation of this energy transition in the near future would maximize the probability for achieving a less than 2 deg C, with a potential 1.5 deg C limit, increase to global temperature over the pre-industrial level by 2100. Our best case scenario utilizes less than 3% of current annual global energy consumption per year with an annual reinvestment of 10% of its growing renewable capacity to make more of itself.

The detrimental impacts of continued fossil fuel consumption particularly with respect to climate change makes a transition to global zero-carbon energy supplies urgent (Hansen et al 2013). Several studies have already demonstrated a path to a global renewable energy (RE) infrastructure in no more than a few decades (Jacobson and Delucchi 2009; Jacobson and Delucchi 2011; Delucchi and Jacobson 2011; Schwartzman and Schwartzman 2011). This transition will require inputs of fossil fuel energy, since the current zero-carbon energy capacity is far from sufficient for the task (Jacobson and Delucchi 2009; Jacobson and Delucchi 2011). Meeting the challenge of energy poverty in the developing world will require the provision of a global level of roughly 3.5 kW per person, corresponding to the minimum required for a world standard high life expectancy (Smil 2008). Here we present the modeling results of a transition to a global wind/solar infrastructure including its climatic impacts. We conclude this transition can be completed in less than 30 years, terminating energy poverty as well as providing additional energy required for climate adaptation and carbon sequestration from the atmosphere to bring down the atmospheric $CO_2$ concentration to safer levels. The implementation of this energy transition in the near future would maximize the probability for achieving a less than 2 deg C, with a potential 1.5 deg C limit (COP21 2015) to global temperature increase over the pre-industrial level by 2100.

In this study, given anticipated global population growth and assumed rapid phase out of high carbon footprint fossil fuels, we calculate global energy usage and concomitant carbon emissions into the atmosphere. Further, using modeling techniques developed by Myhrvold and Caldeira (2012), we estimate the warming expected with these greenhouse forcing contributions. Our analysis clarifies several critical variables, including the following, each determined incrementally over the entire period until the culmination of transition: (a) the amount of energy required (in comparison to current levels and broken down by type); and increments in both (b) global $CO_2$ and methane ($CH_4$) atmospheric concentrations; and, (c) global surface temperatures. Our modeling will hopefully provide an optimal path to be considered by policy makers regarding the implementation of the decarbonation of global energy and its replacement by renewable supplies.

In Schwartzman & Schwartzman (2011), the transition to a future world with 100% renewable energy is modelled using differential equations which allow the computation of the global power capacity of solar energy infrastructure as a function of inputs of fossil fuels and reinvestments of the created renewable energy capacities. Using conservative values of known parameters from existing renewable technologies, in particular, lifespan of photovoltaic panels and wind turbines and the energy return on energy invested (EROI) ratios, they found that very small inputs of current energy consumption per year (~1.0-2.5%) and modest reinvestments of the newly formed solar infrastructure (~10-15%) are sufficient to produce a solar energy infrastructure capable of supplying more than twice modern energy levels on a time scale of 20-30 years. Lifespans of renewable energy technologies were assumed to be 20 to 25 years and the composite EROI ratios for photovoltaic, concentrated solar plants (CSP), and wind farms were taken to be 20:1 to 25:1. Given their already current robust research and development programs, the anticipated improvements in renewable energy technologies will likely make these seemingly dramatic projections quite conservative, as future technologies are implemented.

Myhrvold and Caldeira (2012) model energy transitions, computing warming impacts from carbon emissions for their simulations. However, there is a significant difference between their bootstrapping scenario and our modeling in which the renewable energy capacity exponentially created replaces carbon energy, not leaving it to suddenly disappear at the end (Schwartzman and Schwartzman 2011). Thus, we expected that our solar transition scenarios will commonly result in carbon emissions and resultant warming impacts smaller than the results from their bootstrapping approach. The modeling in previous papers (Schwartzman and Schwartzman 2011, 2013) provided the maximum fossil fuel inputs, hence by inference rough estimates of the maximum global temperature increases, for rapid RE transitions. Here we include in the modeled transitions scenarios with differentiated consumption values of each type of fossil fuel with specified emission factors and compute temperature increases using a climate model. Rapid phase out of coal and other high carbon footprint fuels with conventional liquid oil making up the slack until full wind/solar replacement should have significant future climate consequences because of both the lower C emissions of conventional liquid oil and the complete termination of fossil fuel use in 20-30 years, and this difference informed our modeling.

Starting with the distribution of 2012 energy forms (see Supplemental Table S1) and the UN's best projections for global population growth (WPP 2013), we construct the following seven scenarios: (1) slow initial reductions in fossil fuel energy use in the near-term (fifteen years) followed by rapid reductions (over the next ten years) concomitant with aggressive reinvestment of RE to make build additional RE power capacity; (2) slow initial reductions in high carbon emitting fossil fuel energy use and small increases in lower carbon emitting fossil fuels in the near-term followed by rapid reductions concomitant with less aggressive reinvestments of RE to build additional RE power capacity; (3) similar to (2) in fossil fuel usage but with aggressive reinvestment of RE to make build additional RE power capacity (in the near-term); (4) similar to (1) with slightly slower initial decreases in lower carbon emitting fossil fuels and an even greater reinvestment of available RE resources (for 3 different values of NG greenhouse emissions, resulting in 4a, 4b, 4c, representing lower, middle, and high values of $CH_4$ emission factors, respectively) and more aggressive RE reinvestment for more RE; (5) similar to (3) but with faster ramping down of NG use and more aggressive RE reinvestment for more RE; (6) maintenance of 2012 emissions density levels (both $CO_2$/kWh and kWh/person) as total energy consumption increases according to "business as usual" (with same distribution of energy forms as in 2012); and, (7) maintenance of 2012 emissions density ($CO_2$/kWh) levels with no increase in production (i.e., "no change" for present). Supplemental Table S2 (see supplement) provides the rate of change in use for each fuel type and time interval of the transition; Scenarios 6 and 7 are not included as they do not have preset changes by fuel type. The first four scenarios reduce high carbon emitting fossil fuels much faster than lower carbon emitting fossil fuels. The growth of RE supplies in Scenarios (1) to (5) were computed using different annual growth assumptions (see Supplemental Table S2). While this approach did not use the differential equations of Schwartzman and Schwartzman (2011), the RE levels computed are consistent with those of Schwartzman and Schwartzman (2011) using conservative assumptions regarding input parameters (Table 1). Obviously, climate change concerns motivate the faster initial declines in the high carbon emitting energy forms (see Supplemental Table S1 for the $CO_2$ and $CH_4$ emissions by energy type). However, given some recent findings establishing a much higher level of $CH_4$ emission from natural gas (NG) extraction and use (Howarth 2014; Howarth 2015), scenarios 5a, 5b and 5c differentiate between NG and conventional oil, reducing the former at the same rate as coal and unconventional oil and increasing the latter in the short term, to ensure that enough RE infrastructure is built out early.

Figures 1a-1f shows the output of the seven scenarios, run for twenty-five years into the future, for six important variables: total energy; percentage of energy from energy sources (% RE), power per capita; change in atmospheric $CO_2$ concentrations, change in atmospheric $CH_4$

concentrations, and, temperature change. Table 2 shows only the final values after 25 years in the RE transition. Notice that scenarios 1 and 6 result in equivalent changes in total energy and power per capita but have differences otherwise. The drop in total energy after year 15 (for scenarios 2 and 3) are a result of dramatic declines in fossil fuel use after year 15. Notice that these "dips" are quickly removed once the renewable energy capacity grows to a sufficient scale to replace them. Our optimal scenarios 4a-c and 5a-c arrive at similar total energy levels as scenarios 1, 3 and 6 but do so without the instabilities found in these and usually with a markedly lower $CO_2$ and $CH_4$ concentration change and a lower temperature increase (except in the case of highest $CH_4$ emissions from NG, e.g., 4c and 5c, where $CH_4$ levels can become appreciable after 25 years). In particular, scenarios 4a-b and 5a-b exhibit increases approximately 3-4x smaller than "business-as-usual" and "do nothing," and much smaller than other RE scenarios (other than 1). Not surprisingly then, we observe that these scenarios result in much less warming than the other scenarios as well (~0.26K versus 0.5-0.6K for scenarios 6 and 7 and 0.30-.36K for the other RE scenarios). For 4c and 5c, the two high NG $CH_4$ emission scenarios, we begin to see temperature changes comparable to other RE transition scenarios but much less than "business-as-usual" and "do nothing." In all cases, the benefits of 4a-b and 5a-b (and of the other RE scenarios) start to accrue between 10-15 years after the starting point of introduction of policies to modify fossil fuel use and renewable energy installation. Figure 2 shows the differences between 4a-c and 5a-c in terms of atmospheric $CO_2$ and $CH_4$ concentrations over the 25 year transition to RE. For scenario 4a, the amount of oil and NG required to reach the 93% RE (with total energy 2.1 times larger than 2012 levels) in 25 years is 2,228 EJ for conventional oil and 1,649 EJ for NG, amounting to 18% and 11% respectively of their reserves (Hansen et al 2013).

Note that the $CH_4$ peaks and then declines rapidly during the transition because of its significantly lower atmospheric lifespan than $CO_2$.

Scenario 4a serves as our optimal path to an energy system run almost entirely solar and wind energy that would be self-sustaining; 4b, 5a and 5c scenarios also get us to the same point with a bit more climate change, while 5c provides a more optimal path if $CH_4$ emissions from NG are at the very high end of current estimates. It provides for sufficient power per capita, reaching 90% of the minimum required, 3.5 kW/capita, for world standard high life expectancy (Smil 2008, Schwartzman and Schwartzman 2011) in just 10 years and exceeded it by 30% at the end of the 25 year transition, hence providing additional energy, compared to the present baseline, needed for climate adaptation, and carbon sequestration from the atmosphere to bring down the atmospheric $CO_2$ concentration to safer levels. The actual level of this increment needs study but some preliminary estimates are now available. For example, if 100 billion metric tons of carbon, equivalent to 47 ppm of atmospheric $CO_2$, were industrially sequestered from the atmosphere it would require 5.9 to 18 years of the present global energy delivery (18 TW), assuming an energy requirement of 400 to 1200 KJ/mole $CO_2$ (House *et al.* 2011; Zeman 2007). This requirement would of course be reduced by the use of agriculturally-driven carbon sequestration into the soil. A shift to wind and solar-generated electricity as an energy source could reduce the required power level by 30% once a global system is created (Jacobson and Delucchi 2009; Jacobson *et al*. 2014). And, it should be noted that scenario 4a does not result in a reduction in total energy production at any point in the transition (note that scenario 2 does not meet the criteria for sufficient energy production over the twenty-five years). Scenario 4a reduces future temperature change by more than half, especially when compared to the "business-as-usual" model. It is important to note that our greenhouse gas model is the same as Ricke & Caldeira (2014) and Zhang *et al.* (2014), so the decadal lapse rate in peak warming from $CO_2$ emission found in Ricke and Caldeira (2014) must be contemplated in our case as well. Lastly, humans have the fossil fuel resources today to build the solar/wind infrastructure necessary to make the transition complete. For scenario 4a, 18% of global reserves of conventional oil are used to completely terminate fossil fuel consumption (1). Thus, if implemented, scenario 4a provides for a meaningful means to a world divorced from nearly all fossil fuels and their negative climate implications within a quarter decade.

Already available reliable and cheap storage technologies, along with tapping into geothermal energy, will facilitate the expansion of these renewables (Budischak et al 2013; Jacobson et al 2015; Fairley, 2015). A big enough array of wind turbines, especially offshore, can likely generate a baseload supply without the need to supplement it with separate storage systems (Kempton et al 2010; Archer and Jacobson, 2007). Further, with the progressive expansion of a combined system of wind, photovoltaics and concentrated solar power in deserts a baseload will be created, simply because the wind is blowing and the sun is shining somewhere in the system linked to one grid (e.g., MacDonald et al 2016; Jacobson 2016). Meanwhile baseload would be supplemented by petroleum, with coal phasing out first, on the way to a completely wind/solar global energy infrastructure. The costs for the challenges of intermittency and grid modernization should be absorbed from savings achieved from energy efficiency/conservation and the reduction of health costs corresponding to progressive reduction in air pollution (UNEP Year Book 2014), with a systematic reduction of the subsidies going to fossil fuels, direct and indirect, estimated to be over $5 trillion/year (Coady et al 2015).

A comparison of our current fossil fuel "energy in" (to produce and invest in future fossil fuel) to our anticipated "energy in" to create wind/solar capacity (as determined by current

EROIs, ~20-25, as stated earlier) convinces us that the energy needed for storage and grid modernization is already freed up in the early phases of wind/solar transition, recognizing the role of aggressive energy conservation in buildings, transport and other sectors. Today, we estimate that ~5 to 10% of total fossil fuel consumption goes as "energy in," assuming a conservative EROI of petroleum and coal (10 to 20) (Hall et al 2014), neglecting the ~14% of total energy consumption derived from other sources such as nuclear power and hydropower (BP 2015). In comparison, in our conservative "best case" scenarios (either 4a or 5a), the percentage of required energy "going in" annually to build the large scale renewable capacity (in ~25 years) is ~2-3% of present energy consumption level and a reinvestment of ~10% of wind/solar capacity, both annually. Interestingly, this ~10% of wind/solar capacity which is reinvested annually, which is comparable to today's present energy consumption level reinvestment, is sufficient to make RE capacity sustainable even with a growing population and changes in affluence worldwide.

Is a doubling of present energy consumption in 25 years ending up with a composite of RE technologies a wild stretch of the imagination? We note that wind power alone could supply this level of energy generating capacity several times over (Lu, McElroy and Kiviluoma 2009). Consider the following example, suppose 5 MW capacity wind turbines supply all this energy, with a 35% capacity factor. Then 36 TW, double the present primary energy consumption would require 21 million wind turbines produced in 25 years, assuming the lifespan of this technology exceeds this timespan. We submit that this production is within the technical capacity of the global economy, noting that 90 million cars and commercial vehicles were globally produced in 2014 alone (2014 Product. Stat.).

We demonstrate that the following outcomes are *technically* achievable using current wind/solar power technologies in the next 25 years, if this transition commences in the near future: (1) the virtually complete elimination of anthropogenic carbon emissions to the atmosphere (derived from energy consumption); (2) the capacity for maximizing the probability of achieving a less than 2 deg C, with a potential 1.5 deg C limit to global temperature increase over the pre-industrial level by 2100, taking into account the approximately 0.8 deg C in the pipeline as heat already stored in the ocean (Chen and Tung 2014; Rogelj et al 2013) if anthropogenic $CO_2$ is sequestered from the atmosphere on a continuing basis for roughly 100 years into the future (Cao and Caldeira 2010; Gasser et al 2015); and, (3) the provision of the minimum per capita energy consumption level required to achieve the world standard high life expectancy. There is more than sufficient reserve of the lowest carbon footprint fossil fuel, conventional oil, to make this transition possible. We are not naïve to believe that the formidable political economic obstacles do not exist to implementing this transition. Nevertheless, its chances improved with the news that in 2014 over 100 GW of new wind/solar capacity excluding large hydropower was created (Frankfurt 2015), coupled with the apparent stabilization of global $CO_2$ emissions from the energy and industrial sector in 2014 and 2015 (Jackson et al 2016).

Methods

We programmed the energy scenarios within the context of our solar transition model as follows so as to approximate these outcomes in 25 years: (a) $CO_2$ emissions should drop by more than 90%; (b) energy production should exceed 900 quads (more than 67% higher than 2012 levels) and be over 95% from renewable sources; (c) energy consumption per capita should exceed 3 kW per person as this has been shown to the approximate consumption level required to get people out of energy poverty reaching world standard life expectancy values (Smil 2008; Schwartzman and Schwartzman 2013; increased during the early stages of the transition in order to enable the faster development of RE infrastructure and increased assurances that total power available will be sufficient to provide every human with this level of power over the entire duration of the transition). Note that since nuclear, biomass, biofuels, hydropower, and geothermal power will not be necessary to make the transition (only accounting for 11% of 2012 energy production), consumption of each of them is kept constant at starting levels for the duration of our model (25 years). For each scenario, we estimate the total annual $CO_2$ and $CH_4$ atmospheric contribution (in Tg) over the course of the transition (see Supplemental Table S3).

Energy data from 2012 was used (BP 2013) instead of the most recent data available (2014; BP 2015), but using 2014 data would not change our conclusions. The global primary energy consumption in 2014 was 2.7% greater than for 2012, with less than 1% increase in each fossil fuels fraction of the total consumption in 2014. Our modeling results would be only slightly different, and the same outcome at 25 years would be obtained with a slightly higher phaseout of fossil fuels in the earlier years.

In the estimation of these emissions levels, we made the following assumptions: (a) no $CO_2$ production in operation of solar PV, concentrated PV, wind turbines, or nuclear (IPCC 2007); (b) $CO_2$ emissions from unconventional NG (via hydraulic fracking) is assumed to be equal to conventional NG (though research suggests that it may be significantly higher due to leakage (Caulton et al 2014; Howarth 2014); the latter estimates that they are equal to or greater than coal, for a few decades after consumption); (c) $CO_2$ emissions from modern biomass is assumed to be the same as traditional biomass; and, (d) $CO_2$ emissions from unconventional oil will be comparable to emissions from coal. Details of inputs and assumptions are provided in Supplementary file. Having computed the greenhouse gas contributions to the atmosphere, we estimate the actual ambient concentration of $CO_2$ and $CH_4$ in the atmosphere given these fossil fuel derived inputs using a technique described in the Supplementary file.

Acknowledgements

We would like to thank Ken Caldeira at Stanford University for his support and counsel during this project.

Author contributions

P.D.S. and D.W.S. conceived the research. All authors contributed to the design of the research. P.D.S and D.W.S analysed the data. All authors contributed to writing the manuscript.

Competing financial interests

The authors declare no competing financial interests.

Figure & Table Legend

Figures

1 Output of seven transition scenarios. a. Total Energy Produced (EJ) Globally. b. Percentage of Total Energy that is Renewable Energy (RE). c. Power per capita (global). d. Change in atmospheric $CO_2$ concentration. e. Change in atmospheric $CH_4$ concentration. f. Change in temperature (global).

2 Comparisons of optimal scenarios (4a-c & 5a-c): (a) [$\Delta CH_4$] and (b) [$\Delta T$].

Tables

1 Comparing RE growth in Scenarios (1-3, 4a & 5a) to those of Schwartzman & Schwartzman (2011)
Details the rate of growth in RE used in the model used in this research as compared to that observed in previous research

2 End of Transition Scenario Output Details the output of the model's seven scenarios after 25 years of transition to RE.

(In Supplement)

S1 Constants Used in $CO_2$/$CH_4$ Model
Details the expected release of $CO_2$ and $CH_4$ in the use of various energy sources; values used in the model used here to determine future atmospheric concentrations of these gases.

S2 Model Scenarios, Rates of Growth/Decline per Year
Details the annual rates of change of fossil fuels and RE (renewable energy) used in the model over the lifespan of the model.

S3 Scenario Output (in 5 year increments)
Details the output of the model's seven scenarios with snapshots of annual values each five years during the duration of the transition to nearly 100% RE (or similar percentages of RE as now, in Scenarios 6 and 7)

Supplemental text:

The amount of $CO_2$ and $CH_4$ in the atmosphere at any point in time can be estimated by a convolution of the emissions over time with an impulse response function kernel that describes the atmospheric lifetime of each of the two principal GHGs:

$$m(t) = \int_0^t E(s) \cdot G(t-s) ds$$

The change of GHG in atmospheric concentration is:

C(t) = m(t)/molarmass/molesinatm/fillfactor where molarmass is molar mass of $CO_2$ (44.01 g/mol), or $CH_4$ (16.0426 g/mol). molesinatm is amount of moles in atmosphere, fillfactor is atmospheric to tropospheric abundance, 1 for $CO_2$ and 0.973 for $CH_4$ (Prather, Holmes and Hsu 2012).

The change in atmospheric concentration of a given greenhouse gas depends on many factors, including changes in concentrations of other GHGs and in the climate. Nevertheless, following the IPCC we approximate the change in GHG concentrations by a simple impulse response function (Prather, Holmes and Hsu 2012; Joos et al 2013; IPCC 2013).

$$G_{CO_2}(t) = 0.2173 + 0.2240\, e^{(-t/394.4)} + 0.2824\, e^{(-t/36.54)} + 0.2763\, e^{(-t/4.304)}$$

$$G_{CH_4}(t) = e^{(-t/12.4)}$$

where G(t)'s represent the concentration of $CO_2$ and $CH_4$ respectively, at any given time t after a unit release of each of these gases in the atmosphere at time t = 0. Table S2 (in supplement) provides the incremental atmospheric increases of $CO_2$ and $CH_4$ (in ppm) during the transition.

With these future atmospheric greenhouse gas concentrations in hand, we determine temperature changes expected in the global atmosphere using modelling techniques developed by Myhrvold and Caldeira (2012). We use a simple one-dimensional heat equation with Neumann boundary conditions to estimate the impact on climate of GHG emissions.

$$\frac{\partial \Delta T}{\partial t} = k_v \frac{\partial^2 \Delta T}{\partial z^2}$$

$$\left.\frac{\partial \Delta T}{\partial z}\right|_{z=0} = \left.\frac{\lambda \Delta T - RF(t)}{\rho\, k_v\, f\, c_p}\right|_{z=0}$$

$$\Delta T|_{t=0} = 0$$

$$\left.\frac{\partial \Delta T}{\partial z}\right|_{z=z_{\max}} = 0.$$

where $f = 0.71$ is the fraction of the earth covered by ocean, and, $\rho$ and $c_p$ are the density and heat capacity of seawater, respectively. The maximum depth $z_{\max}$ is chosen as 4,000 meters. RF(t) is radiative forcing. The calculation of radiative forcing follows the IPCC's approach (IPCC 2013). The climate sensitivity parameter ($\lambda$) is 1.051. The ratio of adjusted radiative forcing to the classical radiative forcing derived from the IPCC formula is 0.775. The thermal diffusivity (kv) is $4.24 \times 10^3$ m$^2$/s.

Please note that, "when appropriately calibrated, these simple equations closely follow the global mean temperature results of more complex 3D coupled atmosphere–ocean simulations" (Caldeira and Myhrvold 2012). While the individual contributions of black carbon (warming) and $SO_2$ (cooling) are large, the net climate effect of black carbon and $SO_2$ emissions is small, so these impacts on warming are not included, likewise other trace gases (Zhang, Myhrvold and Caldeira 2014).

Figure 1 Output of seven transition scenarios
a. Total Energy Produced (EJ) Globally. b. Percentage of Total Energy that is Renewable Energy (RE). c. Power per capita (global). d. Change in atmospheric $CO_2$ concentration. e. Change in atmospheric $CH_4$ concentration. f. Change in global temperature.

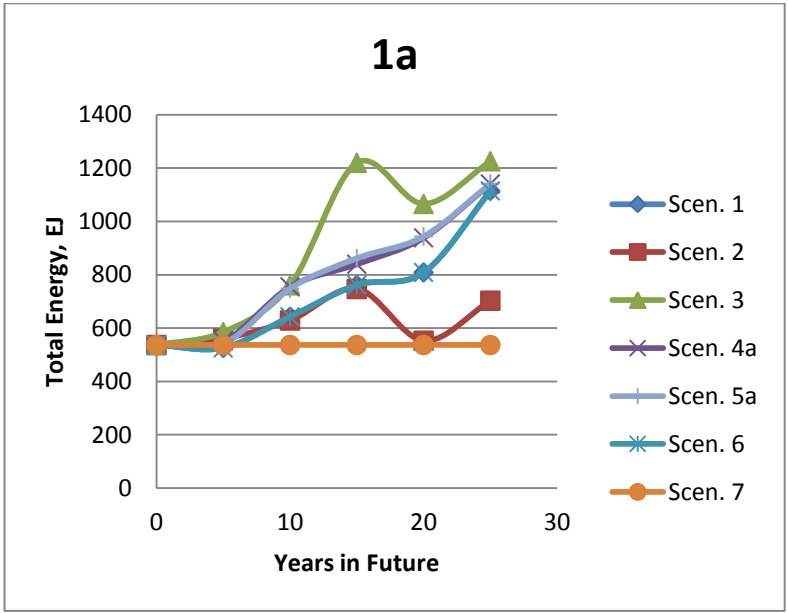
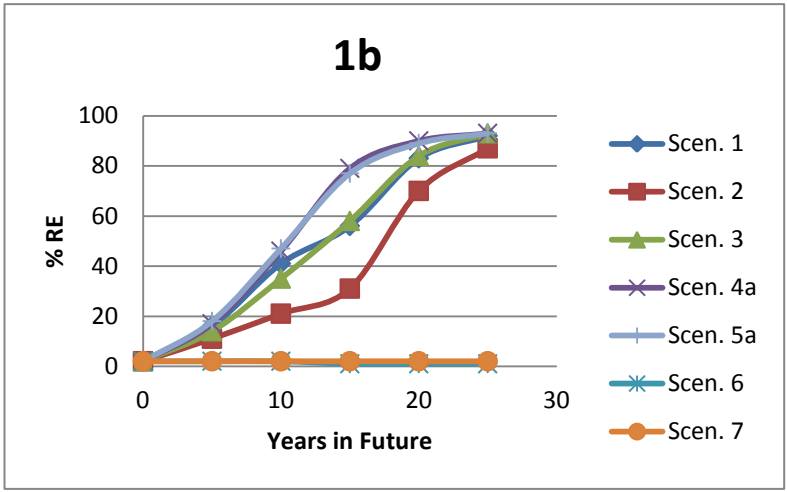

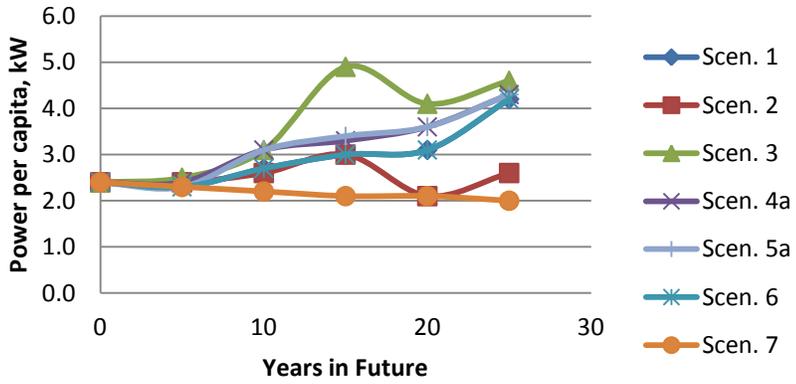

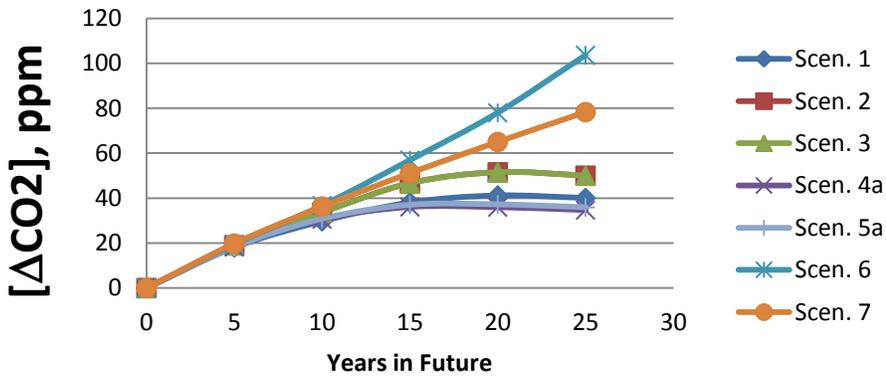

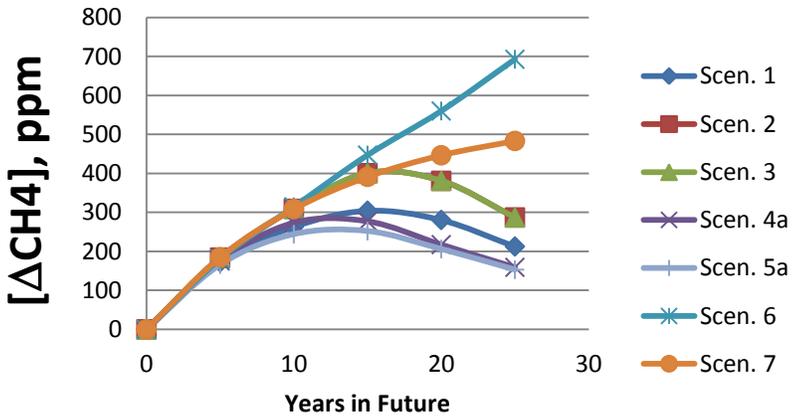

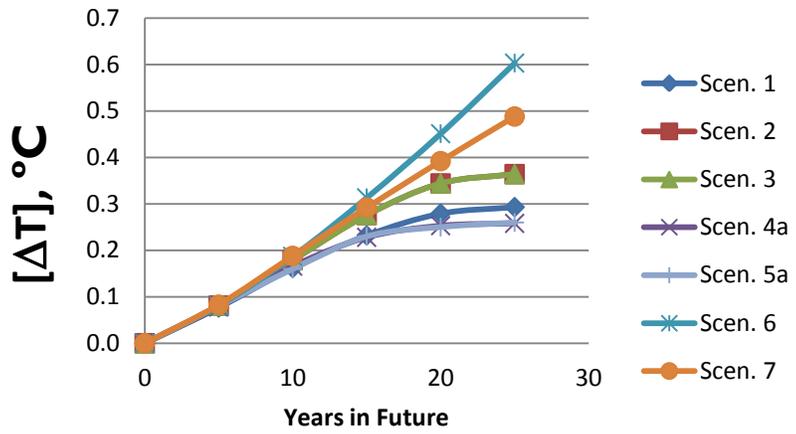

Figure 2  Comparisons of optimal scenarios (4a-c & 5a-c): (a) [ΔCH$_4$] and (b) [ΔT].

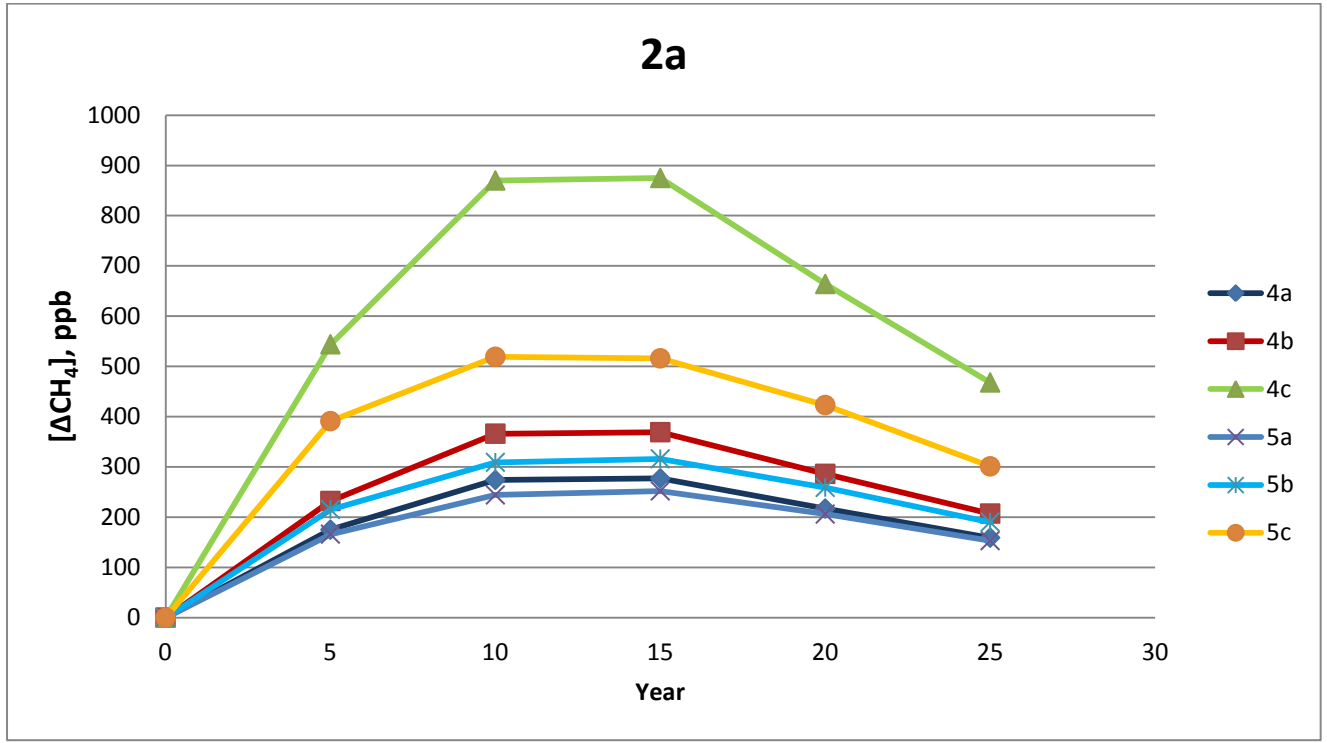

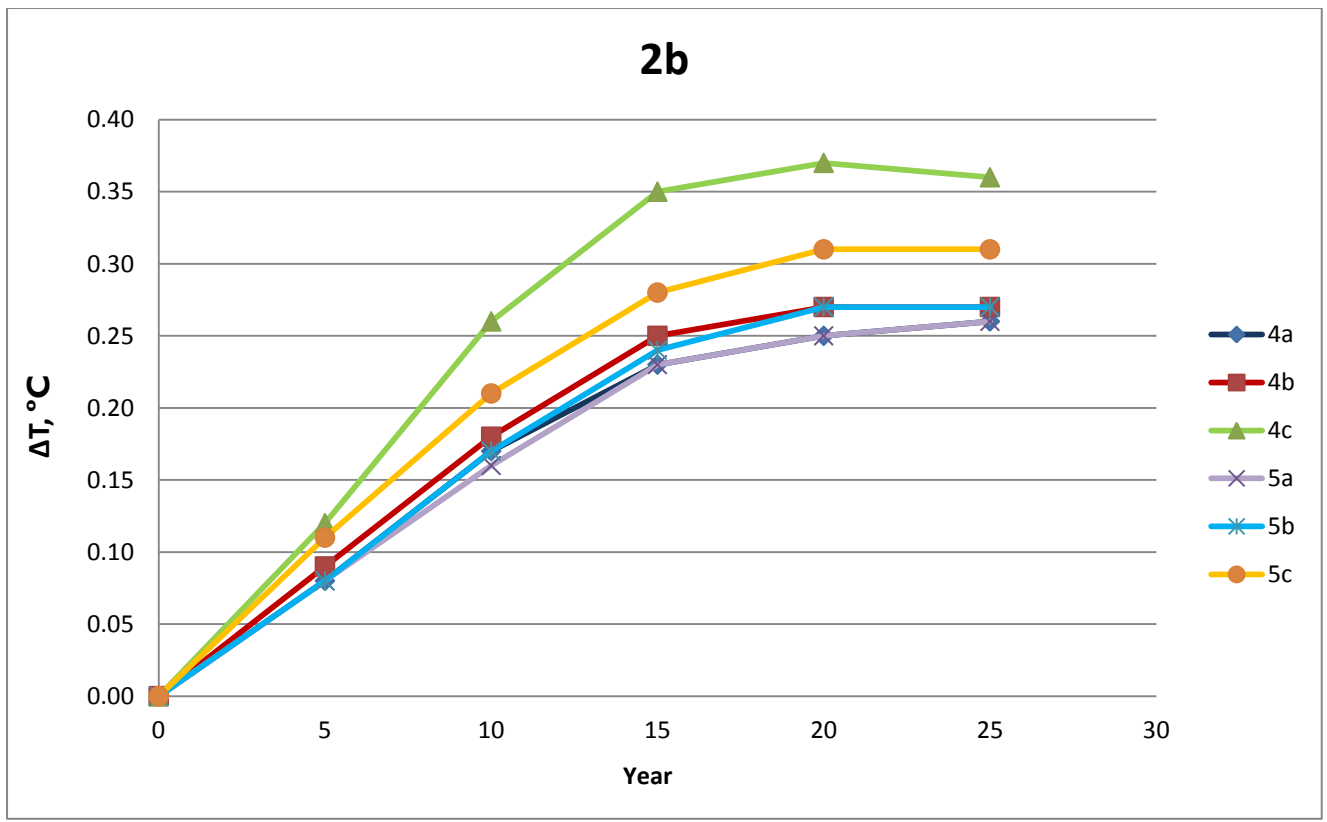

Table 1  Comparisons of RE growth in Scenarios (1-3, 4a & 5a) to those of Schwartzman & Schwartzman (2011)

Scenarios (1-3, 4a and 5a)

| Scenario | Energy Ratio* |
|---|---|
| (1) | 2.08 |
| (2) | 1.31 |
| (3) | 2.28 |
| (4a) | 2.13 |
| (5a) | 2.13 |

Schwartzman & Schwartzman (2011)

| EROI | F | $f_{initial}$ | Energy Ratio** |
|---|---|---|---|
| 25 | 0.10 | 0.020 | 2.2 |
| 20 | 0.15 | 0.020 | 2.5 |
| 20 | 0.11 | 0.028 | 2.0 |

Notes:
(a) EROI = Energy Return On (Energy) Invested
(b) F = percentage of RE capacity that is reinvested annually to build infrastructure of new RE
(c) $f_{initial}$ = percentage of fossil fuel consumption (in 2012) that is reinvested annually to build infrastructure of RE
(d) Energy Ratio = Computed RE (at 25 years)/Total FF energy used (in year 1; 2012)

*This ER is calculated using different scenario rates of the drawdown of fossil fuel consumption and investments in RE capacity.

**This ER is calculated from the solar calculator (found at http://solarutopia.org/solar-calculator/) which uses equations provided in Schwartzman & Schwartzman (2011)

Table 2 Comparing Scenarios at 25 years

| Scenario | Total Energy (EJ) | % RE | Power Avail. (kW/person) | Annual $CO_2$ Emissions (Pg) | Annual $CH_4$ Emissions (Tg) | $[\Delta CO_2]$, ppm | $[\Delta CH_4]$, ppb | $\Delta T$, °C |
|---|---|---|---|---|---|---|---|---|
| 1 | 1,115 | 92 | 4.2 | 2.3 | 10.1 | 40 | 213 | 0.29 |
| 2 | 704 | 87 | 2.6 | 2.8 | 12.1 | 50 | 287 | 0.36 |
| 3 | 1,226 | 93 | 4.6 | 2.8 | 12.1 | 50 | 287 | 0.36 |
| 4a | 1,141 | 93 | 4.3 | 1.7 | 7.0 | 35 | 159 | 0.26 |
| 4b | 1,141 | 93 | 4.3 | 1.7 | 7.5 | 35 | 207 | 0.27 |
| 4c | 1,141 | 93 | 4.3 | 1.7 | 10.1 | 35 | 468 | 0.36 |
| 5a | 1,142 | 93 | 4.3 | 1.8 | 7.4 | 36 | 153 | 0.26 |
| 5b | 1,142 | 93 | 4.3 | 1.8 | 8.0 | 36 | 190 | 0.27 |
| 5c | 1,142 | 93 | 4.3 | 1.8 | 7.0 | 36 | 301 | 0.31 |
| 6 | 1,115 | 1 | 4.2 | 74 | 257 | 104 | 693 | 0.60 |
| 7 | 537 | 2 | 2.0 | 36 | 124 | 78 | 483 | 0.49 |

Note: RE represents "solar" renewable energy (including wind, photovoltaics, and CSP)

Table S1  Constants Used in CO$_2$/CH$_4$ Model

| | 2012 Consumption | CO$_2$ emissions | Sources for CO$_2$ | CH$_4$ emissions | Sources for CH$_4$ |
|---|---|---|---|---|---|
| | EJ/yr | g/MJ | | g/MJ | |
| Coal | 156.7 | 92 | A,E | 0.18 | I1 |
| Natural Gas | 125.5 | 52.4 | A,E | 0.47<br>0.78<br>2.48 | I1<br>I2<br>I2 |
| Oil | | | | | |
|   Oil (conventional) | 169.5 | 76.3 | A,E | 0.18 | I1 |
|   Oil (unconventional--tar sand) | 4.0 | 92 | A,B,E,F | 0.18 | I1 |
| Uranium | 23.5 | 0 | A, G | 0 | H |
| Hydropower | 31.2 | 4.69 | K (C, G) | 0.06 | K |
| Geothermal | 2.5 | 7.14 | C, G | 0 | H |
| Wind | 8.9 | 0 | C, G | 0 | H |
| Solar (electricity) | 3.3 | 0 | | 0 | H |
|   Photovoltaic (PV) | 3.2 | 0 | C, G | 0 | H |
|   Concentrated | 0.1 | 0 | C, G | 0 | H |
| Solar (hot water) | 8.0 | 0 | C, G | 0 | H |
| Biofuels | 3.8 | 49.5 | | 0.01 | J |
|   Biodiesel | 0.8 | 49.5 | C, G | 0.01 | J |
|   Ethanol | 3.0 | 49.5 | C, G | 0.01 | J |
| Biomass | 11.9 | 83.8 | | 0.30 | J |
|   Modern | 2.6 | 83.8 | C, G | 0.30 | J |
|   Traditional | 9.2 | 83.8 | C, G | 0.30 | J |

Sources:

A: Statistical Review of World Energy *BP* June 2013 (Online: http://bp.com/statisticalreview) [consumption multiplied by calorific equivalent]

B: World Energy Outlook 2012 *IEA* p. 104. [Best estimate: 2 mb/day "Extra-heavy oil" (includes Canadian oil sands) = 4.47 EJ]

C: Renewables 2013: Global Status Report *REN 21* (Online: http://www.ren21.net/REN21Activities/GlobalStatusReport.aspx)

D: Hansen, J *et al* 2013 "Assessing 'Dangerous Climate Change': Required Reduction of Carbon Emissions to Protect Young People, Future Generations and Nature *PLOS ONE* 8(12) e81648

E: Climate Change 2007: Mitigation of Climate Change 2007 *IPCC* Cambridge University Press

F: EIA, Voluntary Reporting of Greenhouse Gases Program, updated Jan. 31, 2011 (obtained Feb. 20, 2014)

G: Hodges A W and Rahmani M 2010 Fuel Sources and Carbon Dioxide Emissions by Electric Power Plants in the United States Report FE796 Undated. University of Florida, IFAS Extension.

Note: D&F are background sources.

H: Myhrvold N P and Caldeira K 2012 Greenhouse gases, climate change and the transition from coal to low-carbon electricity *Env. Res. Lett*. 7 014019.

I1: Coal mining, 2013, EPA (Online: http://www.epa.gov/climatechange/Downloads/EPAactivities/MAC_Report_2013-II_Energy.pdf)

I2: Howarth R W 2015 Methane emissions and climatic warming risk from hydraulic fracturing and shale-gas development: implications for policy *Energy and Emission Control Technologies* 3 45-5

J: IPCC 2006 Guidelines for National Greenhouse Gas Inventories (Online: http://www.ipcc-nggip.iges.or.jp/public/2006gl/vol2.html)

K: Role of Alternative Energy Sources: Hydropower Technology Assessment. 2012 National Energy Technology Laboratory  DOE/NETL-2011/1519

Table S2  Model Scenarios, Rates of Growth/Decline per Year

| Scenario | Coal & Oil-u | | NG & Oil-c | | RE | |
|---|---|---|---|---|---|---|
| | First 15 yrs | Next 10 yrs | First 10-15 yrs (in parentheses) | Next 10-15 yrs (in parentheses) | First 10 yrs | Next 15 yrs |
| 1 | -8% | -25% | -2% (15) | -25% (10) | 25% | 10% |
| 2 | -8% | -25% | 2% (15) | -25% (10) | 10% | 10% |
| 3 | -8% | -25% | 2% (15) | -25% (10) | 25%* | 5%* |
| 4a | -8% | -25% | -1% (10) | -25% (10) | 31% | 15%/5%* |
| 4b | -8% | -25% | -1% (10) | -25% (15) | 31% | 15%/5%* |
| 4c | -8% | -25% | -1% (10) | -25% (15) | 31% | 15%/5%* |
| 5a | -8% | -25% | NG: -8% (15); Oil-c: 2% (10) | NG: -25% (10); Oil-c: -25% (15) | 31% | 15%/5%* |
| 5b | -8% | -25% | NG: -8% (15); Oil-c: 2% (10) | NG: -25% (10); Oil-c: -25% (15) | 31% | 15%/5%* |
| 5c | -8% | -25% | NG: -8% (15); Oil-c: 2% (10) | NG: -25% (10); Oil-c: -25% (15) | 31% | 15%/5%* |

Note:

Oil-u represents unconventional oil, such as that obtained via tar sand or shale; Oil-c represents conventionally extracted oil

NG represents Natural Gas

RE represents wind and solar installation

*15% for the next 5 years and 5% for the last 10 years

4a-4c differ only in the $CH_4$ emission value used (0.47 g/MJ, 0.78 g/MJ, 2.48 g/MJ respectively for 4a, 4b and 4c)

5a-5c differ on in the $CH_4$ emission value used (0.47 g/MJ, 0.78 g/MJ, 2.48 g/MJ respectively for 5a, 5b and 5c)

Table S3 Scenario Output

Scenario 1

| Year | Total Energy (EJ) | % RE | Power Avail. (kW/person) | Annual $CO_2$ Emissions (Pg) | Annual $CH_4$ Emissions (Tg) | [$\Delta CO_2$] (ppm) | [$\Delta CH_4$] (ppb) | $\Delta T$ (°C) |
|---|---|---|---|---|---|---|---|---|
| 0 | 537 | 2 | 2.4 | 35.5 | 124 | 0 | 0 | 0.000 |
| 5 | 526 | 16 | 2.3 | 28.6 | 105 | 18 | 173 | 0.077 |
| 10 | 643 | 41 | 2.7 | 23.6 | 91 | 30 | 263 | 0.162 |
| 15 | 762 | 56 | 3.0 | 19.8 | 80 | 38 | 304 | 0.232 |
| 20 | 809 | 83 | 3.1 | 6.0 | 25 | 41 | 281 | 0.279 |
| 25 | 1,115 | 92 | 4.2 | 2.3 | 10 | 40 | 213 | 0.293 |

Scenario 2

| Year | Total Energy (EJ) | % RE | Power Avail. (kW/person) | Annual $CO_2$ Emissions (Pg) | Annual $CH_4$ Emissions (Tg) | [$\Delta CO_2$] (ppm) | [$\Delta CH_4$] (ppb) | $\Delta T$ (°C) |
|---|---|---|---|---|---|---|---|---|
| 0 | 537 | 2 | 2.4 | 35.5 | 124 | 0 | 0 | 0.000 |
| 5 | 561 | 11 | 2.4 | 32.5 | 123 | 19 | 184 | 0.081 |
| 10 | 629 | 21 | 2.6 | 31.4 | 127 | 34 | 309 | 0.179 |
| 15 | 747 | 31 | 3.0 | 31.7 | 134 | 47 | 401 | 0.276 |
| 20 | 553 | 70 | 2.1 | 7.8 | 33 | 52 | 381 | 0.344 |
| 25 | 703 | 87 | 2.6 | 2.8 | 12 | 50 | 287 | 0.364 |

Scenario 3

| Year | Total Energy (EJ) | % RE | Power Avail. (kW/person) | Annual $CO_2$ Emissions (Pg) | Annual $CH_4$ Emissions (Tg) | [$\Delta CO_2$] (ppm) | [$\Delta CH_4$] (ppb) | $\Delta T$ (°C) |
|---|---|---|---|---|---|---|---|---|
| 0 | 537 | 2 | 2.4 | 35.5 | 124 | 0 | 0 | 0.000 |
| 5 | 585 | 14 | 2.5 | 32.5 | 123 | 19 | 184 | 0.081 |
| 10 | 761 | 35 | 3.1 | 31.4 | 127 | 34 | 309 | 0.179 |
| 15 | 1,220 | 58 | 4.9 | 31.7 | 134 | 47 | 401 | 0.276 |
| 20 | 1,066 | 84 | 4.1 | 7.8 | 33 | 52 | 381 | 0.344 |
| 25 | 1,226 | 93 | 4.6 | 2.8 | 12 | 50 | 287 | 0.364 |

Scenario 4a

| | | | | | | | | |
|---|---|---|---|---|---|---|---|---|
| 0 | 537 | 2 | 2.4 | 35.5 | 124 | 0 | 0 | 0.000 |
| 5 | 551 | 17 | 2.4 | 29.5 | 110 | 18 | 175 | 0.078 |
| 10 | 755 | 46 | 3.1 | 25.3 | 99 | 31 | 274 | 0.166 |
| 15 | 840 | 79 | 3.3 | 9.6 | 33 | 36 | 277 | 0.228 |
| 20 | 939 | 90 | 3.6 | 3.2 | 12 | 36 | 217 | 0.252 |
| 25 | 1,141 | 93 | 4.3 | 1.7 | 7 | 35 | 159 | 0.258 |

Scenario 4b

| | | | | | | | | |
|---|---|---|---|---|---|---|---|---|
| 0 | 537 | 2 | 2.4 | 35.5 | 163 | 0 | 0 | 0.000 |
| 5 | 551 | 17 | 2.4 | 29.5 | 147 | 18 | 232 | 0.085 |
| 10 | 755 | 46 | 3.1 | 25.3 | 134 | 31 | 366 | 0.181 |
| 15 | 840 | 79 | 3.3 | 9.6 | 41 | 36 | 369 | 0.248 |
| 20 | 939 | 90 | 3.6 | 3.2 | 14 | 36 | 286 | 0.271 |
| 25 | 1,141 | 93 | 4.3 | 1.7 | 8 | 35 | 207 | 0.274 |

Scenario 4c

| | | | | | | | | |
|---|---|---|---|---|---|---|---|---|
| 0 | 537 | 2 | 2.4 | 35.5 | 376 | 0 | 00 | 0.000 |
| 5 | 551 | 17 | 2.4 | 29.5 | 349 | 18 | 544 | 0.123 |
| 10 | 755 | 46 | 3.1 | 25.3 | 327 | 31 | 870 | 0.258 |
| 15 | 840 | 79 | 3.3 | 9.6 | 87 | 36 | 875 | 0.347 |
| 20 | 939 | 90 | 3.6 | 3.2 | 25 | 36 | 664 | 0.368 |
| 25 | 1,141 | 93 | 4.3 | 1.7 | 10 | 35 | 468 | 0.360 |

Scenario 5a

| | | | | | | | | |
|---|---|---|---|---|---|---|---|---|
| 0 | 537 | 2 | 2.4 | 35.5 | 124 | 0 | 0 | 0.000 |
| 5 | 541 | 18 | 2.3 | 29.6 | 97 | 18 | 166 | 0.077 |
| 10 | 749 | 47 | 3.1 | 26.2 | 81 | 31 | 244 | 0.162 |
| 15 | 862 | 77 | 3.4 | 11.1 | 40 | 37 | 252 | 0.225 |
| 20 | 944 | 89 | 3.6 | 3.5 | 14 | 37 | 206 | 0.253 |
| 25 | 1,142 | 93 | 4.3 | 1.8 | 7 | 36 | 153 | 0.260 |

Scenario 5b

| | | | | | | | | |
|---|---|---|---|---|---|---|---|---|
| 0 | 537 | 2 | 2.4 | 35.5 | 163 | 0 | 0 | 0.000 |
| 5 | 541 | 18 | 2.3 | 29.6 | 123 | 18 | 215 | 0.083 |
| 10 | 749 | 47 | 3.1 | 26.2 | 98 | 31 | 309 | 0.173 |
| 15 | 862 | 77 | 3.4 | 11.1 | 51 | 37 | 316 | 0.239 |
| 20 | 944 | 89 | 3.6 | 3.5 | 16 | 37 | 259 | 0.267 |
| 25 | 1,142 | 93 | 4.3 | 1.8 | 8 | 36 | 190 | 0.273 |

Scenario 5c

| | | | | | | | | |
|---|---|---|---|---|---|---|---|---|
| 0 | 537 | 2 | 2.4 | 35.5 | 124 | 0 | 0.0 | 0.000 |
| 5 | 541 | 18 | 2.3 | 29.6 | 110 | 18 | 391 | 0.106 |
| 10 | 749 | 47 | 3.1 | 26.2 | 99 | 31 | 519 | 0.211 |
| 15 | 862 | 77 | 3.4 | 11.1 | 33 | 37 | 516 | 0.282 |
| 20 | 944 | 89 | 3.6 | 3.5 | 12 | 37 | 423 | 0.310 |
| 25 | 1,142 | 93 | 4.3 | 1.8 | 7 | 36 | 301 | 0.311 |

Scenario 6

| | | | | | | | | |
|---|---|---|---|---|---|---|---|---|
| 0 | 537 | 2 | 2.4 | 35.5 | 124 | 0 | 0 | 0.000 |

| 5  | 526   | 2 | 2.3 | 34.7 | 121 | 19  | 181 | 0.081 |
| 10 | 643   | 2 | 2.7 | 42.5 | 148 | 37  | 314 | 0.187 |
| 15 | 762   | 2 | 3.0 | 50.3 | 176 | 57  | 447 | 0.313 |
| 20 | 809   | 1 | 3.1 | 53.5 | 187 | 78  | 560 | 0.451 |
| 25 | 1,115 | 1 | 4.2 | 73.7 | 257 | 104 | 693 | 0.603 |

Scenario 7

| 0  | 537 | 2 | 2.4 | 35.5 | 124 | 0  | 0   | 0.000 |
| 5  | 537 | 2 | 2.3 | 35.5 | 124 | 20 | 181 | 0.083 |
| 10 | 537 | 2 | 2.2 | 35.5 | 124 | 36 | 309 | 0.188 |
| 15 | 537 | 2 | 2.1 | 35.5 | 124 | 51 | 391 | 0.292 |
| 20 | 537 | 2 | 2.1 | 35.5 | 124 | 65 | 446 | 0.392 |
| 25 | 537 | 2 | 2.0 | 35.5 | 124 | 78 | 483 | 0.488 |

Note: RE represents "solar" renewable energy (including wind, photovoltaics, and CSP)